\documentclass[fleqn,12pt]{wlscirep}

\usepackage{setspace}   
\usepackage{lineno}
\nolinenumbers
\title{Cold gas outflows from the Small Magellanic Cloud traced with ASKAP}

\author[1*]{N.\ M.\ McClure-Griffiths}
\author[1,2]{H.\ D\'{e}nes}
\author[3]{J.\ M.\ Dickey}
\author[4]{S.\ Stanimirovi\'{c}}
\author[5,6]{L.\ Staveley-Smith}
\author[1]{Katherine Jameson}
\author[1]{Enrico Di Teodoro}
\author[7,6]{James R.\ Allison}
\author[2,8]{J.\ D.\ Collier}
\author[2]{A.\ P.\ Chippendale}
\author[9]{T.\ Franzen}
\author[9]{G\"ulay G\"urkan}
\author[9]{G.\ Heald}
\author[9]{A.\ Hotan}
\author[2]{D.\ Kleiner}
\author[2]{K.\ Lee-Waddell}
\author[2]{D.\ McConnell}
\author[5]{A.\ Popping}
\author[5]{Jonghwan Rhee}
\author[9]{C.\ J.\ Riseley}
\author[2]{M.\ A.\ Voronkov}
\author[2]{M.\ Whiting}

\affil[1]{Research School of Astronomy \& Astrophysics, Australian National University, Canberra ACT 2610 Australia}
\affil[2]{CSIRO Astronomy \& Space Science, PO Box 76, Epping NSW 1710 Australia}
\affil[3]{School of Natural Sciences, University of Tasmania, Hobart TAS, Australia}
\affil[4]{University of Wisconsin, Madison WI, USA}
\affil[5]{International Centre for Radio Astronomy Research (ICRAR), University of Western Australia, Crawley, WA 6009, Australia}
\affil[6]{ARC Centre of Excellence for All Sky Astrophysics in 3 Dimensions (ASTRO 3D)}
\affil[7]{Sub-Dept.\ of Astrophysics, Department of Physics, University of Oxford, Denys Wilkinson Building, Keble Rd., Oxford, OX1 3RH, UK} 
\affil[8]{Western Sydney University, Locked Bag 1797, Penrith, NSW 2751, Australia}
\affil[9]{CSIRO Astronomy and Space Science, PO Box 1130, Bentley WA 6102, Australia}

\affil[*]{naomi.mcclure-griffiths@anu.edu.au}

\begin{abstract}
\end{abstract}

\begin{document}
\newcommand{\HI}{H~{\sc i}}
\newcommand{\halpha}{H${\rm \alpha}$}
\newcommand{\MHI}{${\rm M_{\rm HI}}$}
\newcommand{\Msun}{$\rm {M_{\odot}}$}
\newcommand{\arcdeg}{$^{\circ}$}
\newcommand{\kms}{km\,s$^{-1}$}
\newcommand{\askapsoft}{{\sc ASKAPsoft}}
\newcommand{\miriad}{{\sc Miriad}}
\newcommand{\aaps}{Astron. \& Astrophys. Suppl. Ser.}%
\newcommand{\aj}{Astron. J.}%
\newcommand{\apj}{Astrophys. J.}
\newcommand{\actaa}{Acta Astronom.}
\newcommand{\aap}{Astron. \& Astrophys.}
\newcommand{\araa}{Ann. Rev. Astron. Astrophys.}
\newcommand{\mnras}{Mon. Not. R. Astron. Soc.}%
\newcommand{\pasa}{Publ. Astron. Soc. Aust.}%

\flushbottom
\maketitle
\thispagestyle{empty}
\onehalfspacing
{\bf 
Feedback from massive stars plays a critical role in the evolution of the Universe by driving powerful outflows from galaxies that enrich the intergalactic medium and regulate star formation.\cite{hopkins12} An important source of  outflows may be the most numerous galaxies in the Universe: dwarf galaxies.  With small gravitational potential wells,   these galaxies easily lose their star-forming material in the presence of intense stellar feedback.\cite{ferrara00,hopkins12}  Here, we show that the nearby dwarf galaxy, the Small Magellanic Cloud (SMC), has  atomic hydrogen outflows extending at least 2 kiloparsecs (kpc) from the star-forming bar of the galaxy.  The outflows are  cold, $\mathbf{T<400~{\rm K}}$, 
and may have formed during a period of active star formation $\mathbf{25 - 60}$ million years (Myr) ago.\cite{rubele15,hagen17}  The total mass of atomic gas in the outflow is $\mathbf{\sim 10^7}$ solar masses, $\mathbf{{\rm M_{\odot}}}$, or $\mathbf{\sim 3}$\%  of the total atomic gas of the galaxy.   The  inferred mass flux in atomic gas alone, $\mathbf{\dot{M}_{HI}\sim 0.2 - 1.0~{\rm M_{\odot}~yr^{-1}}}$,  is up to an order of magnitude greater than the star formation rate.  We suggest that most of the observed outflow will be stripped from the SMC through its interaction with its companion, the Large Magellanic Cloud (LMC), and the Milky Way, feeding the Magellanic Stream of hydrogen encircling the Milky Way.}  

\begin{figure}[ht]
\centering
\includegraphics[width=\linewidth]{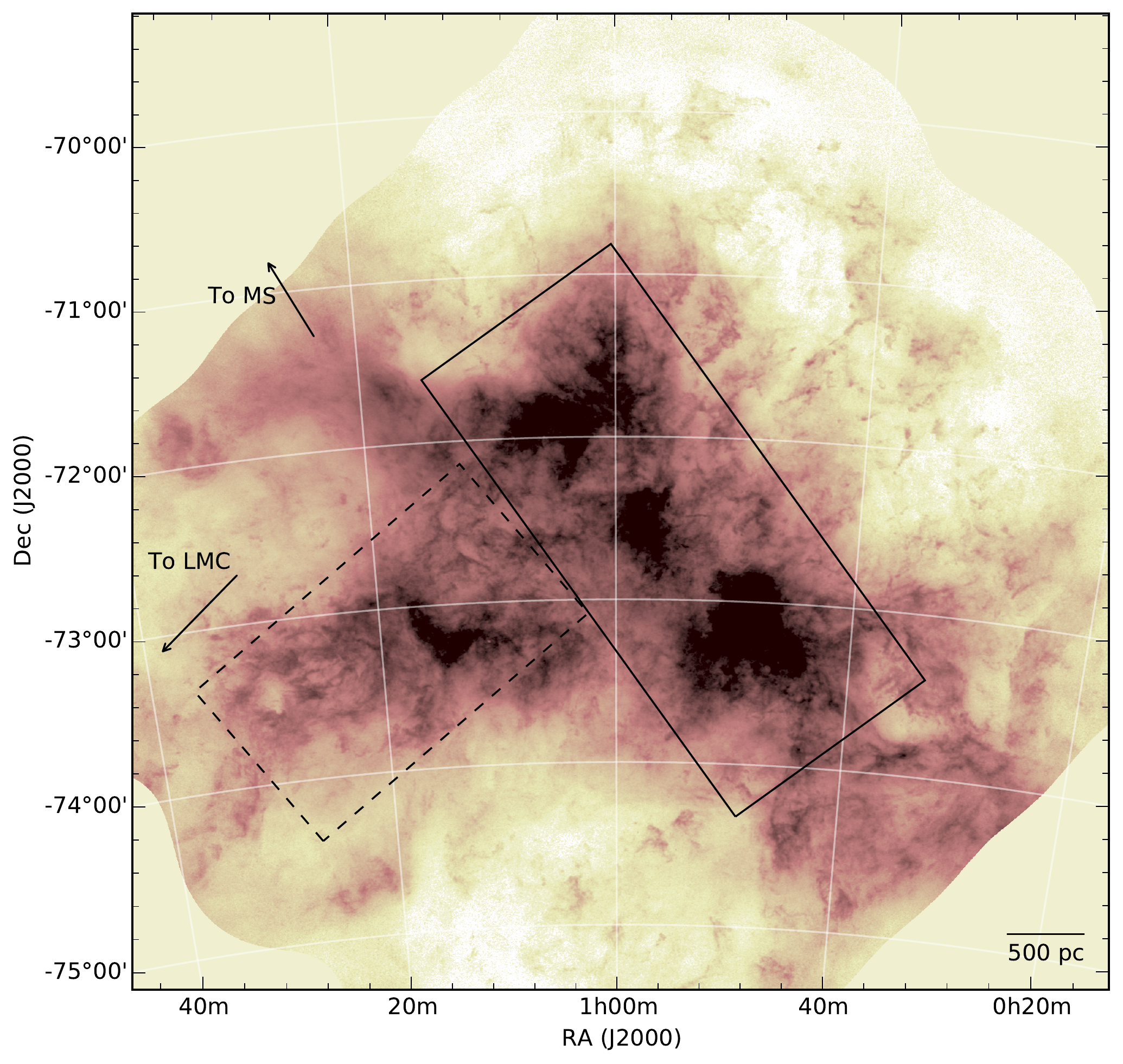}
\caption{Peak \HI\ brightness temperature intensity of the Small Magellanic Cloud from ASKAP and Parkes.  The image has a resolution of about 30 arcsec (8.7 pc) across the whole ASKAP field-of-view, which spans $\sim 5.2$ kpc. The colorscale uses power-law scaling with an exponent of $0.4$ between $3$ and $90$ K. The SMC ``bar`` is indicated by the solid black box and the "wing" by the dashed black box. Arrows indicate the directions to the Large Magellanic Cloud (LMC) and the Magellanic Stream (MS).} 
\label{fig:peak_intensity}
\end{figure}

The SMC, at a distance of 60 kpc,\cite{jacyszyn16}  provides an excellent laboratory to study feedback in an interacting gas-rich dwarf galaxy with sensitivity and resolution that is unattainable almost anywhere else in the Universe.   Using Australian SKA Pathfinder (ASKAP) commissioning data in the 21-cm line of atomic hydrogen (\HI) we have imaged the SMC at $35^{\prime\prime} \times 27^{\prime\prime}$ resolution across a $5.3^{\circ} \times 5^{\circ}$ field-of-view, revealing extensive cold gas outflows from the galaxy.  An image of the peak \HI\ brightness is shown in Figure~\ref{fig:peak_intensity}.  The resolution of this image is the highest ever achieved in \HI\ on this field, revealing features that are ten times smaller in area than previous images.  The gaseous structure of the main body of the SMC is described by two dominant regions as marked on Figure~\ref{fig:peak_intensity}: the so-called bar, extending from the north to the south-west, and containing the majority of the dense gas and star formation and the wing, which extends in the direction of the LMC.\cite{bolatto07}  The bar, as traced by Cepheid variables, is extended along the line-of-sight with the closest part to the north at a distance of $57 - 63$ kpc and containing the younger stars ($<150$ Myr), with the south-western region at $>63$ kpc and older.\cite{jacyszyn16} 

\begin{figure}[ht]
\centering
\includegraphics[width=\linewidth]{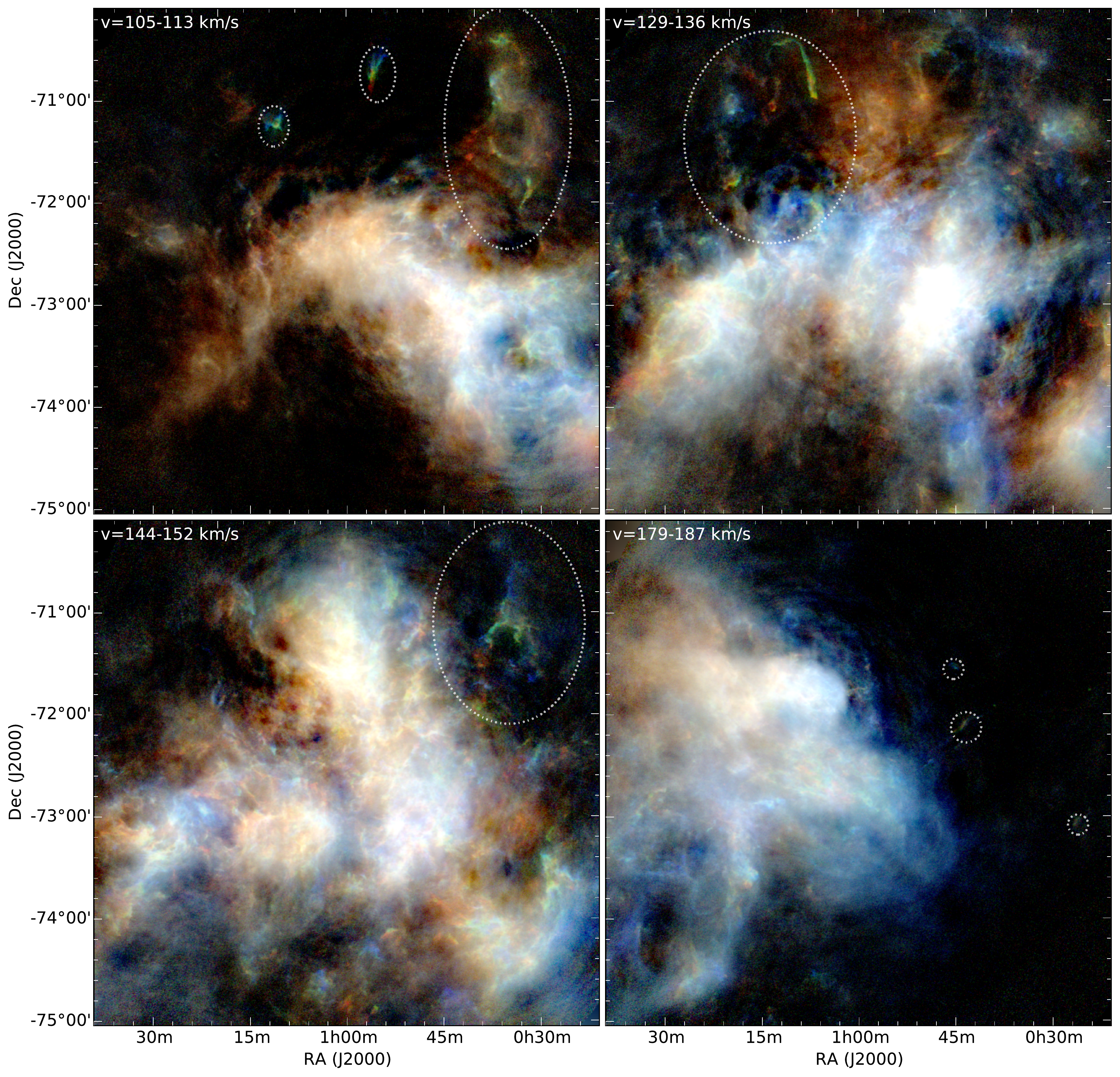}
\caption{ Three-colour images of \HI\ emission for four velocity ranges ($v_{LSR}=105-113~{\rm km~s^{-1}}$, $129 - 136~{\rm km~s^{-1}}$, $144-152~{\rm km~s^{-1}}$, $179-187~{\rm km~s^{-1}}$).  Red, green and blue are assigned sequentially to adjacent velocities, each displayed between $T_{b}=1$ K and $70$ K using a ``arcsinh`` scaling function.  The colorscale of all panels is the same.  Anomalous velocity features are visible as distinctly colored features.  The feature at ($\alpha$, $\delta$) $\approx 00^h49^m$, $-71^{\circ}15^{\prime}$ at $v_{LSR}=105-113~{\rm km~s^{-1}}$ shows a strong velocity gradient of $\sim 20~{\rm km~s^{-1}~kpc^{-1}}$ along its length as evidenced by the red-green-blue transition.  The most massive ( $4.5\times 10^4$ \Msun) single feature is the large encircled structure at $v_{LSR}=105-113~{\rm km~s^{-1}}$.  The green ``hooked`` shape arc to the North-East of the SMC Bar at $01^h06^m$, $-70^{\circ}50^{\prime}$, $v=129 - 136~{\rm km~s^{-1}}$  makes up part of a non-contiguous shell (encircled with a dashed line) that matches with H$\alpha$, emission as shown in Figure~\ref{fig:halpha}.  Three of the ``high-velocity`` cloud examples are seen as faint emission in the western part of the $v=179-187~{\rm km~s^{-1}}$ panel, a region that otherwise appears empty. 
\label{fig:multi-color}}
\end{figure}

Our data have revealed an extensive network of exo-galactic \HI\ features at distances up to 2 kpc from the main body of the SMC.  These features are both spatially and kinematically anomalous.  They can be characterized into three loose categories:   comet-shaped head-tail clouds; enormous looping \HI\ filaments; and compact high-velocity clouds at velocities deviating up to $65$ \kms\ with respect to the rest of the SMC emission.  The majority of these features lie to the North-West and North-East, with comparatively little exo-galactic structure to the South, apart from the well-known SMC Wing.  Some of the most distinctive features are described below and shown in the multi-colour spectral channel images displayed in Figure~\ref{fig:multi-color}.

\begin{figure}[ht]
\centering
\includegraphics[width=\linewidth]{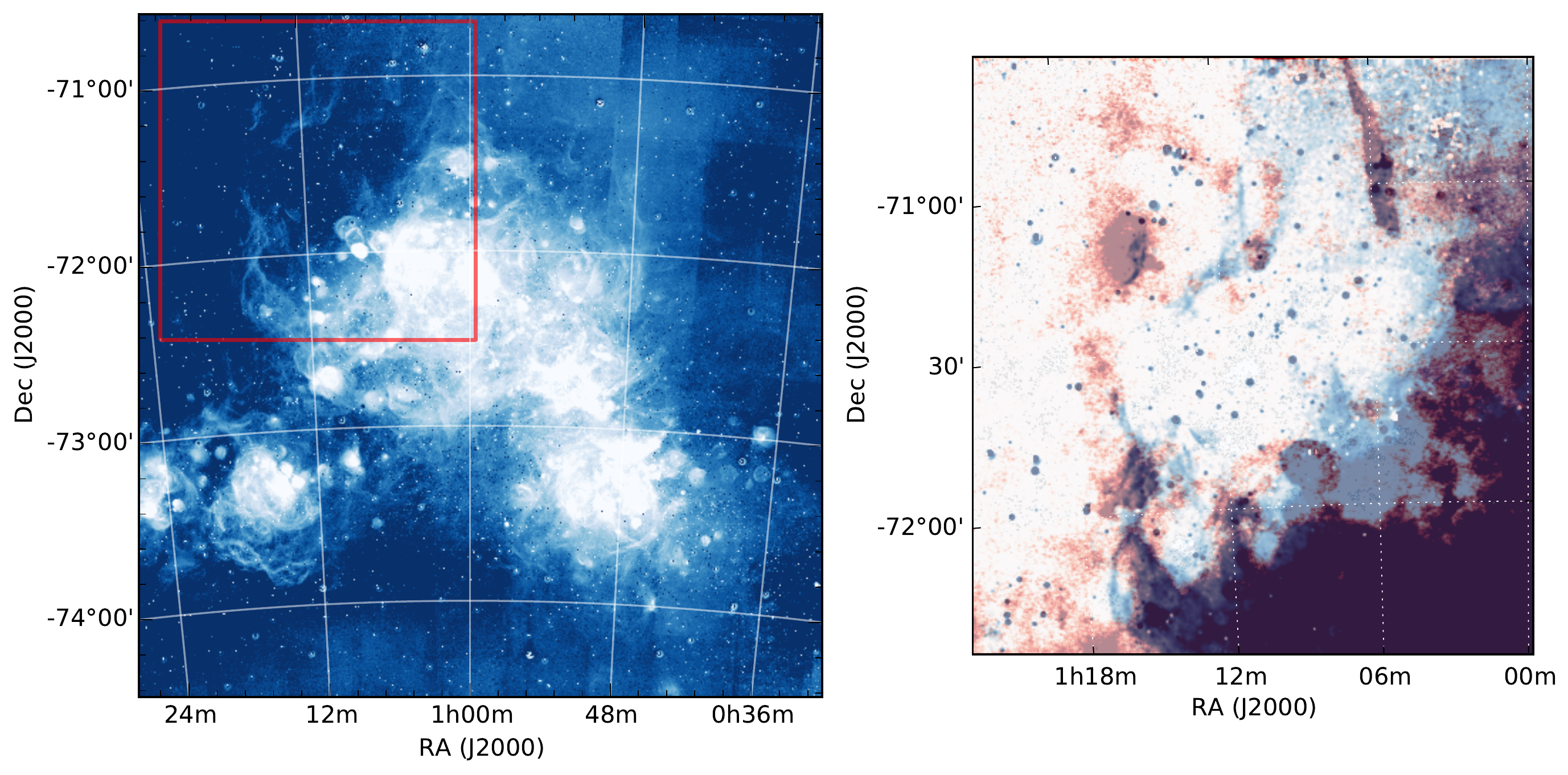}
\caption{ SMC \halpha\ emission from the MCELS survey.  The colour-scale has been saturated to enhance the diffuse emission that extends from the star forming bar.  Features of particular interest are the filamentary structures that are enclosed in the red box and extend up through $(\alpha, \delta) \approx 01^{h}15^{m}$, $-72^{\circ}00^{\prime}$.  These structures are highlighted in the right panel, which shows \HI\ integrated over the velocity range $v_{LSR} = 129 - 136~{\rm km~s^{-1}}$ in dark blue, overlaid with \halpha\ emission in red.  The \halpha\ traces the eastern edge of the non-contiguous \HI\ shell described in Figure~\ref{fig:multi-color}.}
\label{fig:halpha}
\end{figure}

Examples of all three feature categories are shown in Figures~\ref{fig:multi-color} and Figure~\ref{fig:halpha}. Some spatially narrow, comet-shaped structures appear as distinct features to the north of the Bar (Figure~\ref{fig:multi-color}, $v=105-113~{\rm km~s^{-1}}$).  These cometary structures lie between $\sim 1.5$ kpc and $2.6$ kpc  from the bulk of the SMC \HI\ emission at these velocities.  The head-tail structures are oriented radially away from the SMC bar with highly discrepant velocities, typically blue-shifted by  $\sim 50$ \kms\ with respect to the rest of the \HI\ nearest those positions.  Some features show strong velocity gradients of $\sim 20~{\rm km~s^{-1}~kpc^{-1}}$.  These features have plane-of-sky widths, at a distance of 60 kpc, as narrow as $10 - 20$ pc.  Spectra through most of the filaments shown in Figure~\ref{fig:multi-color} show unresolved velocity linewidths ($\Delta v < 3.9$ \kms), indicating temperatures of $T< 400$ K.   Tracing the ridge of the filamentary emission back towards the SMC bar points towards some of the brightest star-formation areas of the SMC Bar (Figure~\ref{fig:halpha}).  The individual narrow, filamentary features shown in Figure~\ref{fig:multi-color} have \HI\ masses in the range $4.2 - 9.4 \times10^4$  \Msun.

Some of the filamentary features appear to be parts of fragmented shells.  Those visible to the north-east of the SMC Bar in Figure~\ref{fig:multi-color} ($v=129-136~{\rm km~s^{-1}}$) form a supershell approximately 2 degrees (2.1 kpc) in  diameter.   To the north-west of the SMC bar there is more extended \HI\  that is spatially and kinematically distinct from the rest of the SMC, visible in Figure~\ref{fig:peak_intensity} \& \ref{fig:multi-color} ($v=105-113~{\rm km~s^{-1}}$ and $v=144-152~{\rm km~s^{-1}}$), centered around ($\alpha$, $\delta$) $\approx~00^h 44^m$, $-71^{\circ}00^{\prime}$.     The top part of the feature is located more than $1.9$ degrees ($2$ kpc) from the peak of the SMC \HI\ emission within the Bar and has a total \HI\ mass of $2.2\times10^6$ \Msun. 

In addition to  features extending perpendicular to the Bar, we observe seven \HI\ clouds at locations behind or in front of the main body of the SMC, with anomalously high velocities.  These clouds have  deviation velocities of $ 35~{\rm km~s^{-1}} < |v_{dev}| < 65~{\rm km~s^{-1}}$, where the deviation velocity is the difference between the cloud velocity and SMC rotational velocity along that sight line.\cite{stanimirovic04} Three examples are seen as faint, compact clouds in Figure~\ref{fig:multi-color} ($v=179-187~{\rm km~s^{-1}}$).   The high-velocity clouds are small ($\sim 10 - 30$ pc) with low peak brightness temperatures of $T\sim 4 - 10$ K.  These clouds have masses in the range $600$ to $4.5 \times10^{4}$  \Msun, with a median mass of $\sim 6000$ \Msun.  Unlike the filamentary structures, all of these clouds have extended velocity widths, on the order of $20 - 30$ \kms.  The compact size of these features at deviation velocities similar to the filaments and with linewidths comparable to the velocity gradients leads us to believe that these could be filaments viewed nearly end-on, such that their velocity gradients are projected as radial velocities.

Anomalous velocity, exo-galactic \HI\ features can have two obvious origins: either outflowing  or accreting gas.  Examples of the former include the  stellar wind and supernovae-driven arches of \HI\ and dust emission that are observed  high above the Milky Way disk\cite{mcgriff06a,pidopryhora07}  and edge-on galaxies such as NGC 891.\cite{howk97,howk00} These can even show head-tail shapes.  Morphologically similar structures are seen as outflow in simulations of feedback in galaxies.\cite{hopkins12}   Accreting gas structures, such as high velocity clouds, can have similar morphology, such as the Smith Cloud, which is known to be falling in to the Milky Way.\cite{lockman08}  Resolving the outflow or infall nature generally relies on either matching the kinematic and geometric properties of the features to infall or outflow models, or clear associations with star formation activity within the galaxy.

In some locations the SMC exo-galactic structures appear to emanate  from within the galaxy.  This supposition is supported in particular with very faint \halpha\ emission, as observed  by the Magellanic Cloud Emission-line Survey (MCELS),\cite{smith99,winkler15} revealing breaking shells of emission and faint tendrils that extend away from the galaxy towards both the North-West and North-East of the Bar (Figure~\ref{fig:halpha}). Some of these faint H$\alpha$ tendrils trace the eastern-most edge of the \HI\ structure described above as a non-continuous \HI\ shell at $v\sim 130~{\rm km~s^{-1}}$ and containing the  "green hook".  This structure is a clear indication of an extended star-formation driven supershell, very much like those observed in the Milky Way and NGC 891.\cite{howk97,howk00}   The $>2$ kpc diameter broken \HI\ + H$\alpha$ shell is similar in morphology with the extensive population of \HI\ supershells observed within the main gas distribution.\cite{staveley-smith97}  Similar morphologies are predicted by simulations of galaxy feedback,\cite{hopkins12} where the gas distribution is dominated by erupting shells and supershells whose column density decreases with increasing distance from the galaxy. 

Further evidence of outflow is in the integrated \HI\ spectrum measured in the full area to the northwest of the SMC bar.  The spectrum shows a red-shifted line-wing extending from LSR velocity $v_{LSR} \sim 190~{\rm km~s^{-1}}$ to  $v_{LSR} \sim 250~{\rm km~s^{-1}}$.  The \HI\ in this wing blends smoothly into the main \HI\ profile, suggesting an origin within the galaxy.   The minimum  escape velocity of the SMC, if it were isolated, is $\mathbf{V}_{esc} \approx \sqrt2 v_{circ} \approx 85$ \kms, where $v_{circ} = 60$ \kms.\cite{stanimirovic04} If the entire outflow were directed along the line-of-sight, then gas at $v_{LSR}>212~{\rm km~s^{-1}}$ would  escape.  Given the SMC Bar is inclined at $i\approx 40^{\circ}$, the total velocity magnitude for gas outflowing perpendicular to the Bar is likely larger than that measured as the radial velocity.

Together these pieces of evidence:  association of some \HI\ features with \halpha\ emission; morphological similarities to previously identified \HI\ shells and supershells within the SMC\cite{staveley-smith97};  extended \HI\ line-wing; and morphological similarity of some features with outflows in simulations of galaxy feedback\cite{hopkins12}  strongly suggest that the exo-galactic \HI\ around the SMC is the result of an outflow.  The total exo-galactic \HI\ mass to the North of the galaxy is  $M = 1.3\times 10^{7}$ \Msun, $\sim 3$\% of the total \HI\ mass of the SMC.\cite{bruens05}  We have restricted our mass measurement to the Northern side of the galaxy to avoid confusion with the Magellanic Bridge to the South, but this means that our mass is a lower limit as we have only measured the mass over 50\% of the volume available.  If we assume that the outflow is roughly isotropic, our measurement likely under-estimates the total exo-galactic mass by a factor of $\sim 2$.

The measured deviation velocities of the filaments and high velocity clouds imply outflow velocities in the range of $35 - 60$ \kms.  Assuming the exo-galactic gas is supershell-related with a radial scale of $1-2$ kpc, this assumption is supported by the measured relationship between shell radius, $R_s$ and  expansion velocities, $V_{exp}$, for SMC \HI\ shells.\cite{hatzidimitriou05} Given  $V_{exp} = 35 - 60$ \kms\ and exo-galactic gas reaching 2 kpc, the timescale for launch (assuming constant velocity) is on the order of $33 - 56$ Myr.  The star formation history of the SMC indicates a recent burst of star formation ($ 0.1~{\rm M_{\odot}~yr^{-1}}$)  $25 - 60$ Myr ago,\cite{rubele15,hagen17} predominantly in the Northern end of the Bar where the \HI\ features are most prevalent.  This is fully consistent with our rough estimate for the age of the \HI\ features.

If the majority of the measured $\sim 1.3 \times 10^{7}$ \Msun\ of \HI\ mass originated in the most recent burst of star formation, $25 - 60$ Myr ago,  the total \HI\ mass flux is  $\dot{M}_{HI}\sim 0.2 - 0.5~{\rm M_{\odot}~yr^{-1}}$.  Values towards the higher end of the range are more likely because the timescale for the formation of features (20 - 35 Myr), together with expectations for the lifetime of exo-galactic \HI, favor mass expulsion in a more recent burst. The \HI\ mass flux is a lower limit because we only measure over a limited field-of-view within a radius of $\sim 2.5$ kpc around the SMC.   Folding in uncertainties related to the limited volume sampled we estimate that the \HI\ mass flux is likely  higher, $\dot{M}_{HI}\sim 0.2 - 1.0~{\rm M_{\odot}~yr^{-1}}$.

In an isolated galaxy some ejected \HI\ mass would return to the galaxy as so-called fountain material unless it exceeds the escape velocity.\cite{ferrara00}  The mass within the extended line-wing (Figure~\ref{fig:spec}) that is strictly beyond the escape velocity is $2.5 \times 10^6$ \Msun, but given outflow at an assumed inclination of $i\approx 40^{\circ}$\cite{stanimirovic04} that increases to  $5 \times 10^6$ \Msun, almost 40\% of the total \HI\ exo-galactic mass.   In addition, the presence of the Milky Way and LMC in close proximity to the SMC  alter the dynamics of the system.  The Magellanic Stream and Leading Arm are  evidence of tidal and/or ram pressure interactions between the LMC/SMC system and the Milky Way.\cite{donghia16}  The tidal radius of the SMC is difficult to determine given the complex interaction with the LMC, uncertainties in the relative masses of the SMC, LMC and Milky Way, and recent revisions in the proper motions of the Magellanic Clouds. \cite{kallivayalil06a,kallivayalil06b,besla07}  Nonetheless, estimates of the SMC tidal radius based on past orbital models\cite{gardiner96} or recent observations of extended stellar structures in the SMC\cite{carrera17} lie in the range $2 - 5$ kpc.  Similarly, the ram pressure produced by the SMC's passage at $v= 302$ \kms\ through the Milky Way halo\cite{kallivayalil06b} should exceed the SMC's effective gravitational pressure (gravitational force per unit area) outside a ram pressure stripping radius, $r_{str}$, of $2 - 4$ kpc.  We observe no evidence of a radial limit for the \HI\ outflow features within our field-of-view.  It therefore seems likely that most of the exo-galactic \HI\ observed  will not return to the SMC as fountain gas.    Both tidal and ram pressure formation models for the Magellanic Stream\cite{besla12,hammer15} have found it difficult to extract sufficient mass from the LMC and SMC to fully account for the mass and shape of the Magellanic Stream and Leading Arm.  The stellar feedback driven outflow described here may help release significant gas quantities from the SMC's gravitational potential so that it can be distributed throughout the Magellanic System.

The ratio of the total mass flux to the total star formation rate, $\dot{M}_{tot}/{\rm SFR}$ gives the mass loss efficiency.  The star formation rate in the SMC during the period 25 - 60 Myr ago when the exo-galactic gas was launched was on order ${\rm SFR} = 0.15~{\rm M_{\odot}~yr^{-1}}$.\cite{rubele15,hagen17}  Therefore, the mass loss efficiency of \HI\ alone is  $\dot{M}_{HI}/{\rm SFR} = 2 - 10$, depending on assumptions about launching timescale and volume filled.  The FIRE simulations of feedback in resolved dwarf galaxies like the SMC  predict a total mass loss efficiency (\HI\ + ionized) that is $\dot{M}_{tot}/{\rm SFR} \sim 10-20$.\cite{hopkins12} Those simulations also predict that atomic gas is a small fraction (from $1 - 50$\% depending on density) of the total outflow mass. If this prediction holds in the SMC, the \HI\ mass flux we measure may be significantly underestimating the total mass flux. The large mass loss efficiency of \HI\ alone suggests that interacting dwarf galaxies are  extremely effective at driving mass loss.  Ionized hydrogen associated with the outflow is much more difficult to trace, but the extended  H$\alpha$ emission\cite{smith99,winkler15}, as well as soft X-ray emission\cite{sturm14}, and O~{\sc VI}\cite{hoopes02} near the SMC all indicate that ionized gas is present in the region North of the bar and suggest that the outflow may be pressure-driven.  Even ignoring the ionized gas outflow, the mass loss rate in \HI\ alone, indicated by the features discovered here, would be enough to completely quench star formation in the SMC in $0.6 - 3$ Gigayears if it were sustained.

\clearpage
\section*{Methods}
\HI\ data were obtained with the Australian SKA Pathfinder (ASKAP) as part of Commissioning and Early Science observations.  This novel interferometer comprises 36 12-meter dishes, each equipped with a phased array feed (PAF) at its focus, which allows it to form receiving beams electronically\cite{deboer09}  across its $\sim 30~{\rm deg^{2}}$ field-of-view.  This turns ASKAP into a powerful mosaicing instrument, capable of covering large areas of the sky very quickly.  

The data discussed here were observed with a 16 antenna sub-array of ASKAP on 2017 November 3, 4, 5.  These data are among the first collected from a 16-antenna ASKAP (antennas: 1, 2, 3, 4, 5, 6, 10, 12, 14, 16, 17, 19, 24, 27, 28, 30).  The array contained baselines between 22 m and 2.3 km.   For these observations 36 beams were formed on each antenna using the maximum signal-to-noise ratio method.\cite{mcconnell16,hotan14} The beams were arranged in a hexagonal grid on the sky, spaced to optimise sensitivity\cite{mcconnell17} and gave each antenna an equivalent field of view of 20 square degrees.  The SMC was observed for 12 hours on each of three successive nights; the beam grid was shifted each night to give a resultant 108-point grid with beams spaced by approximately 0.52 degrees, less than half the beam width and so satisfying the Nyquist criterion for a complete sampling of the sky brightness distribution.\cite{ekers79}  The typical system temperature over efficiency averaged over all beams for these observations was $T_{sys}/\eta \approx75$ K. 

The observations were made over a 240 MHz band divided into 12,960 equal channels of $18.5$ kHz width. The data were calibrated with the ASKAPsoft pipeline\cite{cornwell11} according to the following steps.  First, data from a 20 MHz (1410-1430 MHz) band were selected and duplicated to form continuum and \HI\ datasets.  A bandpass calibration was determined from a separate observation of the primary flux calibrator, PKS B1934-638, and applied to both datasets. We formed a continuum image of the field from the continuum dataset (with HI emission frequencies masked) and used it to self-calibrate the data, over several imaging iterations, and so estimate the variation of antenna gains during the observation.  Calibration of the bandpass and antenna gain variations was performed separately for each of the 36 electronically formed beams.  These gain corrections were applied to the HI dataset.

After the calibration in ASKAPsoft, 200 channels ($3.7$ MHz) were exported into the Miriad data reduction package for imaging. Data on the shortest baselines at low elevations experienced some degree of shadowing and were therefore flagged in Miriad. The independent beams were treated as individual mosaic pointings and imaged using traditional mosaicing techniques \cite{sault96,stanimirovic99}.  The calibrated beams were jointly imaged and deconvolved, allowing maximal recovery of the diffuse emission.  The deconvolved image was restored with a beam of $35^{\prime\prime} \times 27^{\prime\prime}$ and velocities were shifted to the Local Standard of Rest (LSR) frame.  The restored ASKAP data cube was then combined in the Fourier domain with Parkes \HI\ data from HI4PI\cite{hi4pi-collaboration16} using standard Miriad techniques\cite{stanimirovic99} and a flux scale factor of 1.0.  The combination of single dish and interferometric data used here gives sensitivity to all physical scales from the resolution limit of 10 pc to the  5.3 kpc width of the field of view.   The final image cube was converted to brightness temperature units assuming a filled beam and has an rms  per $3.9$ \kms\ spectral channel of $\sigma_{T}=0.75$ K. The new ASKAP data cover approximately 50\% more sky area than the previous \HI\ survey of the SMC\cite{staveley-smith97,stanimirovic99} with a beam area that is 0.1 times smaller.

The total outflow mass was calculated from a masked version of the final data cube.  For each channel in the data cube we used the FloodFill algorithm to mask the data from the location of the peak pixel down to a lower threshold, which we set to $8\sigma_{T}$, where $\sigma _{T}= 0.75$ K as measured in the emission-free velocity channels of the cube.  This step effectively masked the bright emission of the SMC in each velocity channel, while maintaining the outflow features that were spatially and kinematically distinct from the rest of the galaxy.  The total column density of the outflow was then calculated from the zeroth moment of this masked data cube for all pixels above $5\sigma_{T}$ and assuming optically thin emission, resulting in a total mass of $1.3 \times 10^7$ \Msun.

\subsubsection*{Data Availability}
The data that support the plots within this paper and other findings of this study are available from the corresponding author upon reasonable request.



\section*{Acknowledgments}
The Australian SKA Pathfinder is part of the Australia Telescope National Facility which is managed by CSIRO. Operation of ASKAP is funded by the Australian Government with support from the National Collaborative Research Infrastructure Strategy. ASKAP uses the resources of the Pawsey Supercomputing Centre. Establishment of ASKAP, the Murchison Radio-astronomy Observatory and the Pawsey Supercomputing Centre are initiatives of the Australian Government, with support from the Government of Western Australia and the Science and Industry Endowment Fund. We acknowledge the Wajarri Yamatji people as the traditional owners of the Observatory site.  The Magellanic Clouds Emission Line Survey (MCELS) data are provided by R. C. Smith, P. F. Winkler, and S. D. Points. The MCELS project has been supported in part by NSF grants AST-9540747 and AST-0307613, and through the generous support of the Dean B. McLaughlin Fund at the University of Michigan, a bequest from the family of Dr. Dean B. McLaughlin in memory of his lasting impact on Astronomy.  This research made use of Astropy, a community-developed core Python package for Astronomy\cite{astropy18}.  Parts of this research were conducted by the Australian Research Council Centre of Excellence for All Sky Astrophysics in 3 Dimensions (ASTRO 3D), through project number CE170100013.  N.Mc.-G. acknowledges funding from the Australian Research Council via grant FT150100024. We gratefully acknowledge contributions by W.\ Raja \& K.\ Bannister to ASKAP commissioning. 

\section*{Author contributions statement}
N.Mc.-G., J.D., S.S., L.S.-S. developed the idea for the project.  H.D. calibrated the ASKAP data.  N.M.-G. and H.D. produced the ASKAP plus Parkes \HI\ data cube.  J.A., J.C, A.C., T.F, G.H, A.H., D.K., K.L.-W., D.M., A.P., J.R., C.R., M.V. \& M.W.\ are members of the ASKAP Early Science and Commissioning team, with responsibility for delivery of ASKAP data.  N.M.-G. wrote the paper with direct contributions from H.D., J.D., S.S., L.S.-S., K.J., \& E.dT.. All authors reviewed the manuscript. 

\section*{Materials and Correspondence} N.\ M.\ McClure-Griffiths, Australian National University, Canberra ACT 2611 Australia; naomi.mcclure-griffiths@anu.edu.au


\section*{Competing Interests}
The authors declare no competing interests.

\begin{figure}[ht]
\centering
\includegraphics[width=\linewidth]{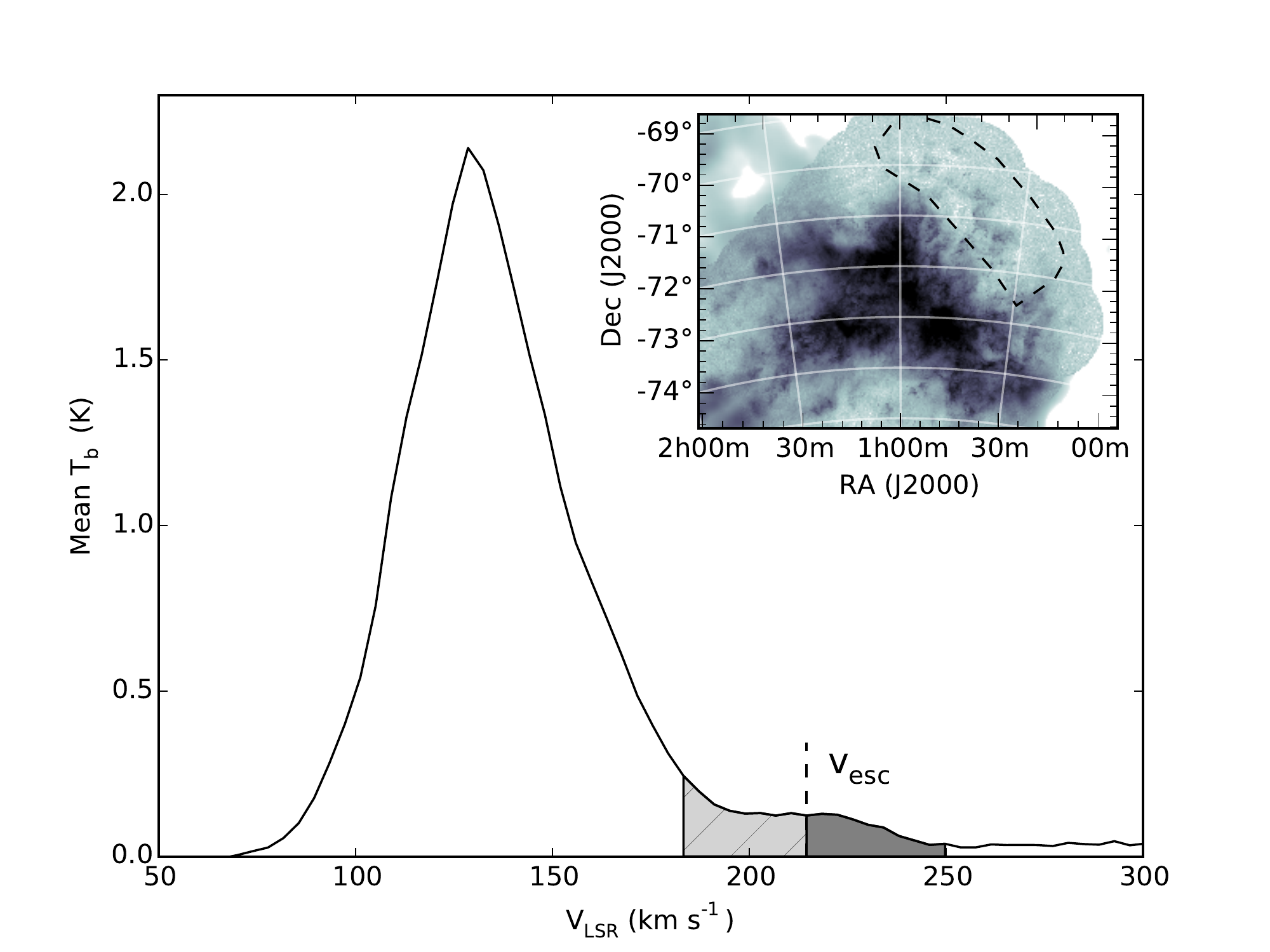}
\caption{ Mean \HI\ spectrum for the region to the north-west of the Bar, displayed as a function of LSR velocity.   The inset shows the SMC peak \HI\ intensity image with the region used for this spectrum indicated with a dotted line.  The  spectrum shows a clear extended line wing from $v_{LSR} \sim 195~{\rm km~s^{-1}}$ to  $v_{LSR} \sim 25~{\rm km~s^{-1}}$.  Gas at velocities beyond $v_{LSR}>212~{\rm km~s^{-1}}$ is beyond the estimated escape velocity for the SMC.  The total mass beyond the escape velocity, indicated by the grey filled area, is on the order of $2.5\times 10^6~{\rm M_{\odot}}$. The grey hatched region shows  gas which would exceed the escape velocity if we assume a galaxy inclination angle of $i=40^{\circ}$, resulting in a total escaped mass of $5\times 10^6~{\rm M_{\odot}}$.}
\label{fig:spec}
\end{figure}

\begin{figure}[ht]
\centering
\includegraphics[width=\linewidth]{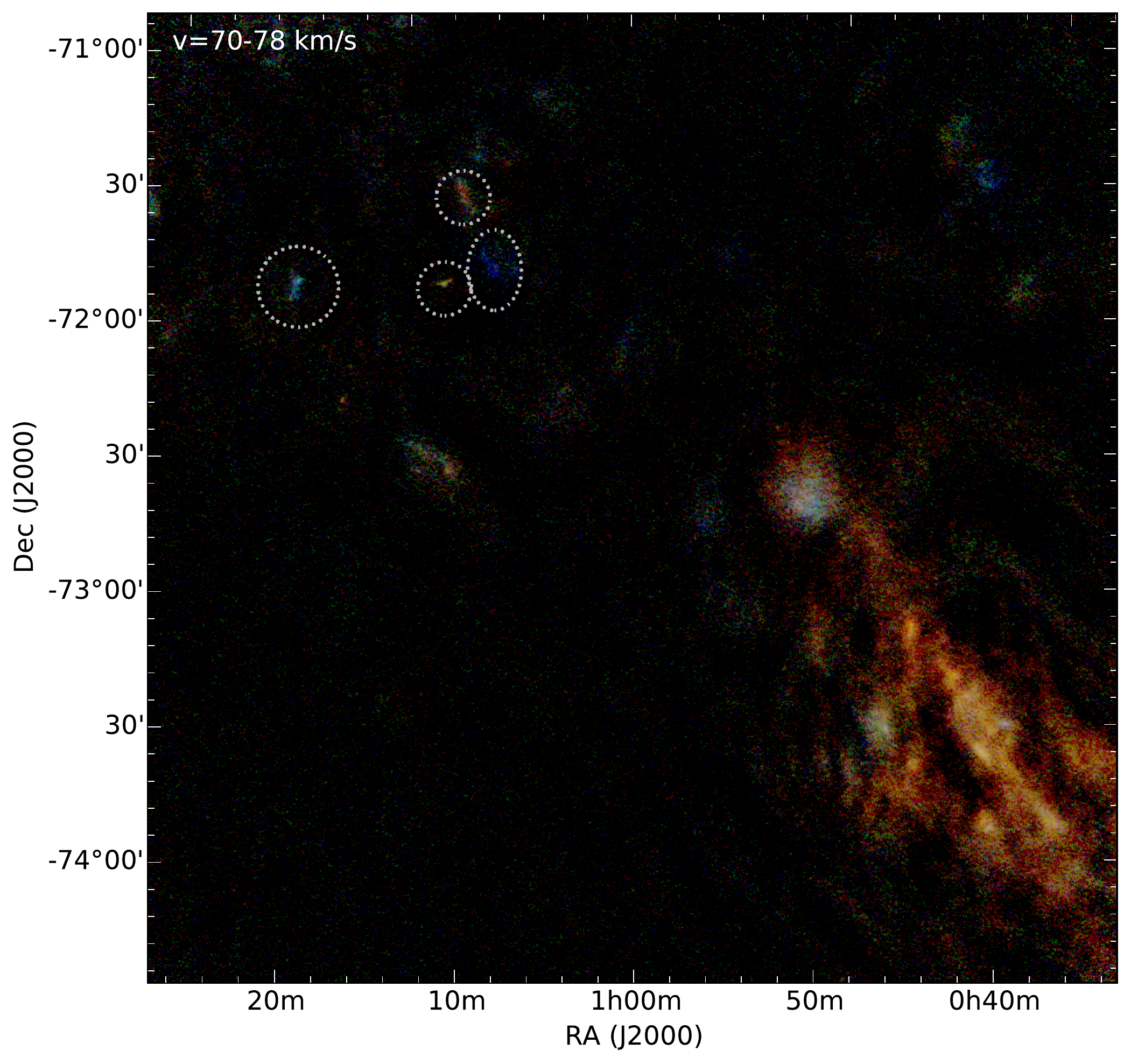}
\caption{{\bf Supplementary Figure}:Four additional ``high-velocity" clouds shown in a three-colour image of \HI\ emission for the velocity range $v_{LSR}=70-78~{\rm km~s^{-1}}$.  Red, green, and blue are assigned sequentially to adjacent velocities, each displayed between $T_{b}=1$ K and $30$ K using a ``arcsinh'' scaling function.  The faint emission of the ``high-velocity'' clouds are circled in the eastern part of the image.  Additional, uncircled \HI\ features that are visible at these velocities connect to larger features at higher LSR velocities or do not exceed a 3$\sigma$ noise threshold for more than one channel. }
\label{fig:spec}
\end{figure}

\end{document}